# Divergent Effects of Factors on Crashes under Autonomous and Conventional Driving Modes Using A Hierarchical Bayesian Approach

**Weixi Ren**, Ph.D. Candidate
Key Laboratory of Road and Traffic Engineering of the Ministry of Education
College of Transportation Engineering, Tongji University
4800 Cao'an Highway, Shanghai, 201804, China
E-mail: 18916225260@163.com

**Bo Yu**, Ph.D., Assistant Professor, Corresponding Author*
Key Laboratory of Road and Traffic Engineering of the Ministry of Education
College of Transportation Engineering, Tongji University
4800 Cao'an Highway, Shanghai, 201804, China
E-mail: boyu@tongji.edu.cn

**Yuren Chen**, Ph.D., Professor
Key Laboratory of Road and Traffic Engineering of the Ministry of Education
College of Transportation Engineering, Tongji University
4800 Cao'an Highway, Shanghai, 201804, China
E-mail: chenyr@tongji.edu.cn

**Kun Gao,** Ph.D., Postdoctoral Research Fellow
Department of Architecture and Civil Engineering, Chalmers University of Technology
Gothenburg SE-412 96, Sweden
Email: gkun@chalmers.se

**Shan Bao**, Ph.D., Associate Professor
Industrial and Manufacturing Systems Engineering Department, University of Michigan-Dearborn, 4901 Evergreen Rd, Dearborn, MI 48128
University of Michigan Transportation Research Institute
2901 Baxter Rd, Ann Arbor, MI, USA, 48109-2150
E-mail: shanbao@umich.edu

**Abstract**

Influencing factors on crashes involved with autonomous vehicles (AVs) have been paid increasing attention. However, there is a lack of comparative analyses of those factors between AVs and human-driven vehicles. To fill this research gap, the study aims to explore the divergent effects of factors on crashes under autonomous and conventional driving modes. This study obtained 180 publicly available autonomous vehicle crash data (96 for the autonomous driving mode and 84 for the conventional driving mode), and 39 explanatory variables were extracted from three categories, including environment, roads, and vehicles. Then, a hierarchical Bayesian approach was applied to analyze the impacting factors on crash severity (i.e., injury or no injury) and type (i.e., rear-end or not) under both driving modes with considering unobserved heterogeneities. The results showed that some influencing factors affected both driving modes, but their degrees were different. For example, daily visitors' flowrate had a greater impact on the crash severity under the conventional driving mode, while the presence of turning movement led to a larger decrease in the likelihood of rear-end crashes in the autonomous driving mode. More influencing factors only had significant impacts on one of the driving modes. For example, in the autonomous driving mode, two sidewalks increased the severity of crashes, while daytime had the opposite effects. As for crash type, on-street parking was positively associated with rear-end crashes of autonomous driving, whereas the presence of trees was negatively associated with that. Similarly, some factors only significantly affected the likelihood of the crash severity (e.g., the number of lanes at crash site) or crash type (e.g., slop) under the conventional driving mode. This study could contribute to the understanding and development of autonomous driving systems and the better coordination and complementarity between autonomous driving and conventional driving.

# 1. Introduction

It is widely believed that full automation driving will be an important direction in the development of transportation engineering and will provide a potential solution to transportation-related issues in safety, efficiency, and mobility (De Winter et al., 2014; Fagnant and Kockelman, 2015; Johnson and Walker, 2016). However, due to the incompleteness and cost of the current technological progress, the perception, identification, and decision-making systems of autonomous vehicles (AVs) are not perfect so far. They cannot effectively deal with all kinds of factors that affect driving safety. Heretofore, there have been hundreds of crashes with the autopilot system turned on, which leads to heavy personal and property losses (National Transportation Safety Board, 2016).

Impacting factors on the crash type and severity of AVs may be different from human-driven vehicles since autonomous driving is integrated and systematically based on big data and artificial intelligence, while conventional driving is personalized (Kockelman et al., 2016). Additionally, there are a series of new problems brought up by AVs. For example, when driving autonomous vehicles on roadways, drivers may use their travel time to accomplish leisure activities which will inhibit their anticipation of possible driving activities and eventually result in a volatile traffic environment. (Gucwa, 2014; Barnard and Lai, 2010; Carsten et al., 2012). Therefore, it is meaningful and imperative to explore the divergent effects of factors affecting safety for autonomous and conventional driving using crash data.

The crash data involved with AVs become more and more available for the public, due to the fact that the regulatory requirements for the development and testing of AVs are gradually relaxed. AVs were allowed to be tested on roadways in September of 2014 (State of California Department of Motor Vehicles, 2019). Companies and manufacturers that were approved to test AVs on California public roads must submit a Traffic Collision Involving an Autonomous Vehicle Report (OL 316) of the full description of the collision and other valid information (California Department of Motor Vehicles (California DMV), 2021).

Crash analyses of AVs have arisen within the past years. Some studies have focused on the factors contributing to AV crash severity levels. The positive association between travel speed and crash severity has been widely reported (Rosen et al., 2011). A substantially higher likelihood of AV-involved injury crashes at intersections was found (Lee and Abdel-Aty, 2005; Retting and Kyrychenko, 2002). The lengthy time for drivers to repossess the authority of the vehicle might increase the likelihood of serious incidents (Merat et al., 2014). Down-slopes, nighttime, involvement of multiple vehicles, and high-density traffic would also increase the likelihood of high crash severity of AVs (Zhang and Xu, 2021). In addition, location at an intersection, presence of roadside parking has been found to be the main positive contributing factors to the severity of AV crashes, while the one-way road would decrease the crash severity. (Xu et al., 2019).

It has been found to be the most frequent AV crash type, representing around 40% of the total number in California during September 2014–May 2019 (Das and Tsapakis, 2020). Many studies have investigated the impacting factors on AV crash types. The probability of an AV-involved rear-end collision increased on one-way streets or in mixed land-use settings (Boggs et al., 2020). When the automated driving system was engaged, AV rear-end crashes were more likely to happen, and AVs were more inclined to be rear-ended by conventional vehicles (Favarò et al., 2017). A study also reported that crashes which occurred under the connected and autonomous vehicle (CAV) driving mode were more likely to be rear-end crashes, and the movement of turning was associated with an increase in the risks of rear-end crashes (Xu et al., 2019). Mixed land use, one-way indicator, and poor lane markings were found to be positively related to rear-end crashes, while the count of public/private schools within a quarter-mile, vehicle year, and so on had a negative impact on this kind of crashes (Boggs et al., 2020). Using rear-end crash data, a statistical analysis by Liu et al. (2021) examined the differences of pre-crash scenarios between AVs and conventional vehicles. The perception-reaction time between AVs and human drivers was found to be different, which was an important cause for the AV crashes.

A lot of research on disengagements has been done based on Disengagement Reports (OL311R) from the California Department of Motor Vehicles (DMV). Causes and contributing factors of disengagements were investigated, and lacking certain numbers of radar and LiDAR sensors installed on AVs were found to significantly induce an AV disengagement (Wang and Li, 2019). Besides, influence factors between disengagements with a crash, disengagements with no crash, and no disengagement with a crash in a mixed traffic environment were also discussed. Variables related to AV systems (such as software failures) and other roadway participants may increase the propensity of disengagement without a crash (Khattak et al., 2020). With the analysis of AV's interactions with other road users before a collision in a temporal manner, the results showed that the most representative pattern in AV crashes was "collision following AV stop" (Song et al., 2021). Unlike these previous studies, this study will not analyze disengagements separately but consider disengagement-related variables as additional explanatory variables to obtain more insights into the pre-crash behavior of AVs, and then distinct effects of factors on crashes between autonomous and conventional driving modes will be further explored in this study.

To analyze the influencing factors of crashes, many methods have been employed in previous studies, such as probit model, binomial/multinomial logistic regression, classification tree, and so on (Haleem and Abdel-Aty, 2010; Xu et al., 2019; Dadvar and Ahmed, 2021). However, reliable and unbiased correlations between crashes and influencing factors cannot be established because of the presence of unobserved heterogeneity. (Mannering and Bhat, 2014; Yu et al., 2021; Khattak and Wali, 2017). The hierarchical Bayesian approach can solve such problems. In addition, since AV crash data are difficult to collect but gradually available, the hierarchical Bayesian approach can use any engineering experiences or justified previous findings as prior knowledge to update the model (Huang et al., 2008; Bakhshi et al., 2021). This approach could also well handle missing data that occur commonly in crash records, by considering the information contained in other observed data (Faes et al., 2011; Ma and Chen, 2018). In addition, such a technique performs well in the estimation of discrete outcome models with smaller sample sizes (Ahmed et al., 2018). For example, this method was applied to analyze correlations of influencing factors of AV-involved crashes, with a sample of 113 available crashes (Boggs et al., 2020).

Given the above, there has been much research on the influencing factors of crashes and disengagements involved with autonomous driving, but there is a lack of comparative analyses of those factors between AVs and human-driven vehicles. To fill this research gap, this study utilizes the publicly available AV crash data in real driving environments and employs the hierarchical Bayesian approach to further explore and examine the differences of impacting factors between autonomous and conventional driving modes from the aspects of both crash severity (i.e., injury or no injury) and types (i.e., rear-end or not). This study will assist the understanding, development, and testing of autonomous driving systems. In the stage of human-machine co-driving, it can also contribute to the better coordination and complementarity of autonomous driving with conventional driving.

## 2. Methodology
### 2.1 Data preparation

As mentioned before, when AVs were involved in a crash while driving on public roads in California, a description of how the collision occurred and other associated factors would be submitted in the Traffic Collision Involving an Autonomous Vehicle Report (OL 316). AVs in the conventionally human-driven mode still need to submit crash reports, therefore, crashes in both driving modes (i.e., the autonomous driving mode and conventional driving mode) were included in this database. These publicly available reports can be downloaded from the website (https://www.dmv.ca.gov/portal/vehicle-industry-services/autonomous- vehicles/ autonomous-vehicle-collision-reports/). By the time of writing this paper, reports from May 2018 to March 2021 were open, fully informative, and available. The Society of Automotive Engineers (SAE) defines six levels of driving automation to describe the full range of driving automation features, from Level 0 (No automation) to Level 5 (full automation) (Taxonomy, 2018). Vehicles in this AV crash database are considered to be conditional automation (Level 3), also known as driver-initiated automation (Boggs et al., 2020; Banks and Stanton, 2016). Level 3 AVs are equipped with the ADAS technologies, sensors, and actuators, capable of automated highway driving, automated city driving, automated valet parking, and evasive maneuvers, but it is still essential for test drivers to take over driving promptly if there is a foreseen crash (Khayyam et al., 2020).

To consider the transitions from AV systems to test drivers, more information from Disengagement Reports (OL 311R) provided by California DMV (https://www.dmv.ca.gov/portal/dmv/detail/vr/autonomous/testing) was added, and these reports consisted of all instances of disengagements occurring when AVs were tested. Then, the disengagement-related data was linked with the AV crash database (i.e., Traffic Collision Involving an Autonomous Vehicle Reports (OL 316)). Since not all disengagements led to a crash, information from Disengagement Reports (OL311R) were matched to the AV crashes that involved disengagements by carefully comparing the specific dates, manufacturer, and vehicle types. In addition, a few crashes involving disengagements which could not be found in Disengagement Reports (OL 311R) but recorded by the descriptions of crashes in OL 316 were manually marked as "the presence of disengagement". In this study, the autonomous driving mode includes two situations: (1) The AV system remained engaged throughout the crash; (2) The driver took over the AV before the crash (i.e., disengagement occurred). Conventional mode indicates that manual mode is employed before the crash for a considerable period and the human driver independently responds to the crash. Crashes in the conventional driving mode were filtered by two criteria: (1) It was emphasized in OL 316 that the vehicle was manually driven before the crash and disengagement was not mentioned; (2) The crash cannot be found in the Disengagement Reports (OL311R). A total of 180 crashes in San Francisco were extracted and used in the final analysis, including 96 crashes in the autonomous driving mode and another 84 in the conventional driving mode. Crashes with disengagement accounts for about 35% in autonomous driving mode.

Six new variables (i.e., disengagement, initiator of disengagement, unwanted behavior of other roadway participants, unwanted movement of AVs, changing lane, deceleration) from Disengagement Reports (OL311R) were fully contained in the dataset for further analyses. Specifically, disengagement reflects the presence or absence of disengagement in the autonomous driving mode. Initiator of disengagement indicates whether the disengagement is initiated by the system or the test driver. The other four variables were extracted from the description of Disengagement Reports (OL311R), which may be the causes of disengagements. Unwanted behavior of other roadway participants means reckless action of another vehicle or another non-vehicle roadway participant, such as a cyclist driving aggressively. Illegal behavior of AVs, such as entering the opposite lane suddenly, is reflected by the unwanted movement of AVs. Changing lanes indicate lane-changing maneuvers of AVs for reasons such as unstable target lane model. Deceleration refers to AVs dropping the speed for safety precaution or other reasons. As for those disengagements only found in the crash reports of OL316, their disengagement-related variables were manually extracted based on the descriptions of crashes.

To better understand the impact of environmental, road, and vehicle characteristics on the safety under different driving modes, this study obtained more explanatory variables through TransBASE: Linking Transportation Systems to Our Health (http://transbasesf.org/transbase/) and Google Earth (https://www.google.com/earth) and then made hard efforts to manually link them to the crash sites through the location of each crash. TransBASE is a free and open online database that currently includes over 200 spatially referenced variables from multiple agencies and across a range of geographic scales, including infrastructure, transportation, zoning, sociodemographic, and collision data, all linked to an intersection or street segment. It is currently used by San Francisco Municipal Transportation Agency. Seven environmental variables (metro stop, land use, muni line, daily visitors' flowrate (DVF), pavement markings conditions, schools, parks) and 13 road variables (street classification, one-way, divided median, marked centerline, bike lane, on-street parking, off-street parking, traffic calming, sidewalk, driveway, crash lanes, speed limit, slope) were obtained from TransBase. Google Earth was used to supplement some information, such as the specific width of the road.

Crash severity and type were chosen to be dependent variables to create a picture of AV crashes. Considering the small amount of data, crash severity was divided into two levels. Crashes with injuries were considered more serious, while a crash without injury (i.e., property damage only) was thought to be of lower severity. The research on crashes with injuries or not had unique implications, especially for autonomous vehicles. The public had great concerns about the safety of the AVs and crashes with injury or death of people have proven to be a potential deterrent to the acceptance and credibility of AVs. In addition, insurance, legal, ethical, economic, and other fields were also interested in whether there was an injury in a crash involved with AVs. In the analysis of crash types, crashes were classified into two groups by whether it was a rear-end crash.

Before modeling, some typical variables that describe driving conditions were picked out and a percent-stacking bar chart was plotted. As can be seen from Figure 1, the proportion of each selected variable, such as the time of crashes (Night), road characteristics (Speed limit, Street width, Number of driveways, Street type), type of the crash places (Daily visitors' flowrate (DVF), Intersection, Land use), and vehicle state at the time of crashes (Turning movement, Vehicle state) was similar, which meant crashes in the autonomous driving mode and conventional driving mode occurred under similar conditions, and they were comparable.

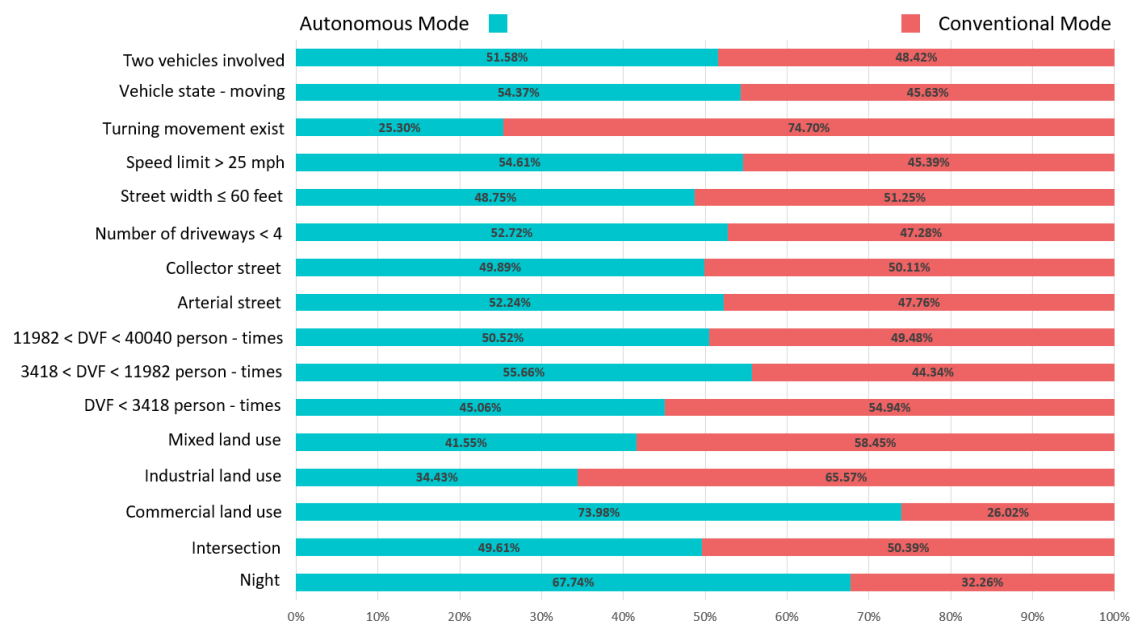

**Fig. 1. A comparison of driving conditions in both driving modes**

Various discrete and continuous variables were obtained. The continuous variables and their descriptive statistics are provided in Table 1. Before the model establishment, this study divided continuous variables into discrete variables. Variables, including the count of public and private schools within a quarter-mile, the count of parks within a quarter-mile, and the count of driveways along the segment, were divided into two groups, according to whether the number was less than 4. The rest continuous variables were split into two groups, such as the number of lanes at the crash site (great than 2 or not), the width of the street in feet (more than 60 feet or not), the speed limit (more than 25mph or not), the slope of the road (larger than 3% or not), etc.

**Tab. 1. Descriptive statistics of continuous variables**

| Variable category | Description | Mean | S.D. | Min | Max |
|---|---|---|---|---|---|
| **Environmental Variables** | | | | | |
| Schools | Count of public and private schools within a quarter-mile | 1.94805 | 1.8499 | 0 | 9 |
| Parks | Count of parks within a quarter-mile | 1.89286 | 1.46636 | 0 | 6 |
| **Road Variables** | | | | | |
| Driveway | Count of driveways along segment | 3.12987 | 1.36579 | 1 | 8 |
| Crash lanes | Number of lanes at crash site | 2.12338 | 1.07453 | 1 | 6 |
| Street width | Width of street in feet | 51.85065 | 19.63816 | 22 | 140 |
| Speed limit | Speed limit of roadway in mph | 25.42208 | 1.71034 | 15 | 30 |
| Slope | Slope in percentage of roadway | 3.41558 | 2.95931 | 1 | 10 |

After discretization of continuous variables, 41 discrete variables used in the model were finally obtained. Besides two dependent variables injury and crash type, Table 2 presents all the 39 explanatory variables, deriving from three categories, including environment, roads, and vehicles. Specifically, there were 14 environmental variables (e.g., trees, land use, weather, roadway surface, etc.), 14 road variables (e.g., bike lanes, street width, number of driveways, etc.), and 11 vehicle variables, (e.g., vehicle damage, turning movement, manufacturer, vehicle year, vehicle state, etc.). The detailed descriptions, distributions, and sources of them are provided.

Tab. 2. Descriptions, distribution, and sources of explanatory variables

| Variable Category | Description | Variable | Autonomous | | Conventional | | Source |
|---|---|---|---|---|---|---|---|
| | | | Num | Percent | Num | Percent | |
| Injury | Someone injured | No* | 74 | 77.08% | 70 | 83.33% | OL 316 |
| | | Yes | 22 | 22.92% | 14 | 16.67% | |
| Crash type | Type of the crash | Rear-end | 57 | 59.38% | 34 | 40.48% | OL 316 |
| | | Other* | 39 | 40.63% | 50 | 59.52% | |
| **Environmental Variables** | | | | | | | |
| Time of day | Time of the crash | Daytime | 60 | 62.50% | 69 | 82.14% | OL 316 |
| | | Night* | 36 | 37.50% | 15 | 17.86% | |
| Involved in the crash | Non-motor vehicles or pedestrians involved in the crash | No | 78 | 81.25% | 58 | 69.05% | OL 316 |
| | | Yes* | 18 | 18.75% | 26 | 30.95% | |
| Intersection | Crash happened at an intersection | No* | 33 | 34.38% | 28 | 33.33% | OL 316 |
| | | Yes | 63 | 65.63% | 56 | 66.67% | |
| Light | Presence of light | Dark* | 54 | 56.25% | 5 | 5.95% | OL 316 |
| | | Daylight | 42 | 43.75% | 79 | 94.05% | |
| Roadway surface | Condition of roadway surface | Dry | 91 | 94.79% | 75 | 89.29% | OL 316 |
| | | Wet* | 3 | 3.13% | 6 | 7.14% | |
| | | Unknown | 2 | 2.08% | 3 | 3.57% | |
| Metro stop | Presence of metro stop | Absence* | 51 | 53.13% | 39 | 46.43% | TransBASE |
| | | Presence | 45 | 46.88% | 45 | 53.57% | |
| Trees | Presence of trees | Absence* | 19 | 19.79% | 23 | 27.38% | TransBASE |
| | | Presence | 77 | 80.21% | 61 | 72.62% | |
| Land use | Land use of the location | Commercial | 26 | 27.08% | 8 | 9.52% | TransBASE |
| | | Industrial | 3 | 3.13% | 5 | 5.95% | |
| | | Mixed or public | 39 | 40.63% | 48 | 57.14% | |
| | | Residential* | 28 | 29.17% | 23 | 27.38% | |
| Weather | Weather at the time of the crash | Clear weather* | 85 | 88.54% | 74 | 88.10% | OL 316 |
| | | Cloudy | 5 | 5.21% | 7 | 8.33% | |
| | | Fog/Visibility | 2 | 2.08% | 0 | 0.00% | |
| | | Raining | 3 | 3.13% | 3 | 3.57% | |
| | | Unknown | 1 | 1.04% | 0 | 0.00% | |
| Muni line | Presence of muni line | Absence* | 20 | 20.83% | 12 | 14.29% | TransBASE |
| | | Presence | 76 | 79.17% | 72 | 85.71% | |
| Daily visitors' | Level of DVF | DVF＜3418 person-times | 30 | 31.25% | 32 | 38.10% | TransBASE |

| Variable Category | Description | Variable | Autonomous | | Conventional | | Source |
|---|---|---|---|---|---|---|---|
| | | | Num | Percent | Num | Percent | |
| flowrate (DVF) | | 3418 person-times≤DVF＜11982 person-times | 33 | 34.38% | 23 | 27.38% | |
| | | 11982 person-times≤DVF＜40040 person-times | 28 | 29.17% | 24 | 28.57% | |
| | | DVF≥40040 person-times* | 5 | 5.21% | 5 | 5.95% | |
| Pavement markings conditions | conditions of pavement markings | Poor* | 6 | 6.25% | 6 | 7.14% | Google Earth |
| | | Adequate | 90 | 93.75% | 78 | 92.86% | |
| Schools | Count of public and private schools within a quarter-mile | Count of schools＞4 | 20 | 20.83% | 16 | 19.05% | TransBASE |
| | | Count of schools≤4* | 76 | 79.17% | 68 | 80.95% | |
| Parks | Count of parks within a quarter-mile | Count of parks＞4 | 6 | 6.25% | 5 | 5.95% | TransBASE |
| | | Count of parks≤4* | 90 | 93.75% | 79 | 94.05% | |
| **Road Variables** | | | | | | | |
| Street classification | Classification of street | High | 1 | 1.04% | 0 | 0.00% | TransBASE |
| | | Arterial | 20 | 20.83% | 16 | 19.05% | |
| | | Collector | 33 | 34.38% | 29 | 34.52% | |
| | | Residential* | 42 | 43.75% | 39 | 46.43% | |
| One-way | One-way street | No* | 62 | 64.58% | 56 | 66.67% | TransBASE |
| | | Yes | 34 | 35.42% | 28 | 33.33% | |
| Divided median | Presence of divided median | Absence* | 80 | 83.33% | 76 | 90.48% | TransBASE |
| | | Presence | 16 | 16.67% | 8 | 9.52% | |
| Marked centerline | Presence of marked centerline | Absence* | 56 | 58.33% | 43 | 51.19% | TransBASE |
| | | Presence | 40 | 41.67% | 41 | 48.81% | |
| Bike lane | Presence of bike lane | Absence* | 70 | 72.92% | 54 | 64.29% | TransBASE |
| | | Presence | 26 | 27.08% | 30 | 35.71% | |
| On-street parking | Presence of on-street parking | Absence* | 15 | 15.63% | 11 | 13.10% | TransBASE |
| | | Presence | 81 | 84.38% | 73 | 86.90% | |
| Off-street parking | Presence of off-street parking | Absence* | 1 | 1.04% | 3 | 3.57% | TransBASE |
| | | Presence | 95 | 98.96% | 81 | 96.43% | |
| | | Absence* | 69 | 71.88% | 58 | 69.05% | TransBASE |

| Variable Category | Description | Variable | Autonomous | | Conventional | | Source |
|---|---|---|---|---|---|---|---|
| | | | Num | Percent | Num | Percent | |
| Traffic calming | Presence of traffic calming device | Presence | 27 | 28.13% | 26 | 30.95% | |
| Sidewalk | Presence of sidewalk | Absence or one-side of segment* | 5 | 5.21% | 7 | 8.33% | TransBASE |
| | | Both sides of segment | 91 | 94.79% | 77 | 91.67% | |
| Driveway | Count of driveways along segment | driveways≥4* | 31 | 32.29% | 33 | 39.29% | TransBASE |
| | | driveways<4 | 65 | 67.71% | 51 | 60.71% | |
| Crash lanes | Number of lanes at crash site | Crash lanes>2 | 36 | 37.50% | 27 | 32.14% | TransBASE |
| | | Crash lanes≤2* | 60 | 62.50% | 57 | 67.86% | |
| Street width | Width of street in feet | Street width>60 feet | 21 | 21.88% | 15 | 17.86% | Google Earth |
| | | Street width≤60 feet* | 75 | 78.13% | 69 | 82.14% | |
| Speed limit | Speed limit of roadway in mph | Speed limit>25 mph | 11 | 11.46% | 8 | 9.52% | TransBASE |
| | | Speed limit≤25mph* | 85 | 88.54% | 76 | 90.48% | |
| Slope | Slope in percentage of roadway | Slope>3% | 42 | 43.75% | 31 | 36.90% | TransBASE |
| | | Slope≤3%* | 54 | 56.25% | 53 | 63.10% | |
| **Vehicle Variables** | | | | | | | |
| Turning movement | Turning movement of the AV | No* | 84 | 87.50% | 53 | 63.10% | OL 316 |
| | | Yes | 12 | 12.50% | 31 | 36.90% | |
| Manufacturer | Manufacturer of the AV | Aurora Innovation, Inc. | 0 | 0.00% | 1 | 1.19% | |
| | | GM Cruise LLC | 79 | 82.29% | 53 | 63.10% | OL 316 |
| | | Lyft, Inc. | 0 | 0.00% | 2 | 2.38% | |
| | | Waymo LLC | 8 | 8.33% | 9 | 10.71% | |
| | | Zoox,lnc | 9 | 9.38% | 19 | 22.62% | |
| Vehicle year | Production year of the AV | 2016 | 9 | 9.38% | 17 | 20.24% | |
| | | 2017 | 20 | 20.83% | 16 | 19.05% | |
| | | 2018 | 0 | 0.00% | 1 | 1.19% | OL 316 |
| | | 2019 | 21 | 21.88% | 15 | 17.86% | |
| | | 2020 | 45 | 46.88% | 34 | 40.48% | |
| | | 2021 | 1 | 1.04% | 1 | 1.19% | |

| Variable Category | Description | Variable | Autonomous | | Conventional | | Source |
|---|---|---|---|---|---|---|---|
| | | | Num | Percent | Num | Percent | |
| Vehicle state | State of AV | Stopped* | 32 | 33.33% | 37 | 44.05% | OL 316 |
| | | Moving | 64 | 66.67% | 47 | 55.95% | |
| Number of vehicles involved | Number of vehicles involved in the crash | 1* | 11 | 11.46% | 13 | 15.48% | OL 316 |
| | | 2 | 84 | 87.50% | 69 | 82.14% | |
| | | 3 | 1 | 1.04% | 2 | 2.38% | |
| Disengagement | Presence of disengagement | Absence* | 60 | 62.50% | 84 | 100.00% | OL 316 &OL311R |
| | | Presence | 36 | 37.50% | 0 | 0.00% | |
| Initiator of disengagement | Initiator of disengagement (system or the test driver) | AV system | 1 | 1.04% | 0 | 0.00% | OL 316 &OL311R |
| | | Test driver | 35 | 36.46% | 0 | 0.00% | |
| | | No | 60 | 62.50% | 84 | 100.00% | |
| Unwanted behavior of other roadway participants | Presence of unwanted behavior of other roadway participants | Absence* | 77 | 80.21% | 84 | 100.00% | OL 316 &OL311R |
| | | Presence | 19 | 19.79% | 0 | 0.00% | |
| Unwanted movement of AVs | Presence of unwanted behavior of AVs | Absence* | 95 | 98.96% | 84 | 100.00% | OL 316 &OL311R |
| | | Presence | 1 | 1.04% | 0 | 0.00% | |
| Changing lanes | Presence of AV's changing lanes | Absence* | 64 | 66.67% | 84 | 100.00% | OL 316 &OL311R |
| | | Presence | 32 | 33.33% | 0 | 0.00% | |
| Deceleration | Presence of AV's deceleration | Absence* | 76 | 79.17% | 84 | 100.00% | OL 316 &OL311R |
| | | Presence | 20 | 20.83% | 0 | 0.00% | |

*Note: * denotes the reference group*

Before modeling, multicollinearity was checked by calculating the variance inflation factors (VIFs) for all independent variables. VIF indicates the extent to which an indicator's variance is captured by the remaining indicators of a given construct and VIF＞10 denotes severe multicollinearity (Assemi and Hickman, 2018; Pan et al., 2019). In this study, the VIF values for all the selected independent variables were less than 10, indicating that the problem of multicollinearity did not exist or could be negligible while modeling.

**2.2 Hierarchical Bayesian approach**

This study applied the hierarchical Bayesian approach to explore the differences of impacting factors on crashes for both autonomous and conventional driving modes while considering the unobserved heterogeneities caused by vehicle companies and vehicle years. This approach was composed of two parts: the hierarchical model and Bayesian inference.

**2.2.1 Hierarchical Model**

The hierarchical method can properly model the potential heterogeneities

(Mannering et al., 2016; Yu et al., 2019b; Bao et al., 2020), so the crash effects of explanatory variables can be analyzed more accurately by using this multi-level structure. In this part, a hierarchical logistic regression model was used to analyze the impact of different influencing factors on crash types and crash severity. In particular, a "Vehicle company & year" unit was considered as a cluster, and there were several sub-clusters per cluster, i.e., each crash.

Previous studies show that taking vehicle units as observation units may reveal crash propensity variation among different vehicles (Chand and Dixit, 2018; Wali et al., 2018). Recently, more companies are becoming permit holders to test their AVs on roadways. The perception recognition system, decision-making system, software algorithm, and computing ability of AVs produced by different companies have a lot of differences (Dadvar and Ahmed, 2021). It should also be noted that autonomous driving technologies are persistently and rapidly advancing, and the vehicle year can reflect the "older" or "newer" technology to a certain extent. The complex influence of such unobserved factors on the correlation between other observed variables and dependent variables, called unobserved heterogeneity, may result in biased indications. These variables cannot be obtained, but they could be reflected by AVs' company and production year to a certain degree. Therefore, this study took vehicles with the same vehicle year from the same company (i.e., the "Vehicle company & year" unit) as an observation unit to alleviate the effects of unobserved heterogeneity.

In the analysis of crash severity, the response variable $Y_{Sij}$ for the $i$th crash in the $j$th vehicle unit is a dichotomous variable, such that $Y_{Sij}=1$ means high severity (i.e., injury crash), while $Y_{Sij}=0$ represents low severity (i.e., no injury crash). As for the analysis of crash type, $Y_{Tij}$ was the response variable, which also takes one of two values: $y_{Tij}=1$ means rear-end crashes, while $Y_{Tij}=0$ represents other types. When describing the structure of the hierarchical Bayesian model, there was no difference in formulas between the analysis of types and severity. To make the expression of equations more concise, $Y_{ij}$ was used to represent $Y_{Sij}$ or $Y_{Tij}$ below. The probability of $Y_{ij} = 1$ is denoted by $\pi_{ij} = \Pr(Y_{ij} = 1)$ which follows a binomial distribution. In level 1(crash level), the probability of $Y_{ij}=1$ is described as follows:

$$logit(\pi_{ij}) = \log\left(\frac{\pi_{ij}}{1-\pi_{ij}}\right) = \beta_{0j} + \sum_{p=1}^{P} \beta_{pj} X_{pij} + \varepsilon_{ij} \quad (1)$$

where $\beta_{0j}$ is the level1 intercept; $\beta_{pj}$ is the regression coefficient for $X_{pij}$; $X_{pij}$ is the value of the $p$th independent variable for crash $i$ for vehicle unit $j$; $P$ is the number of independent variables in level1; $\varepsilon_{ij}$ is the disturbance term with mean zero and variance to be estimated.

In the context of the hierarchical model, the within-crash correlation is specified in the "Vehicle company & year" level (level 2) as:

$$\beta_{0j} = \gamma_{00} + \sum_{q=1}^{Q} \gamma_{0q} Z_{qj} + \mu_{0j} \quad (2)$$

$$\beta_{pj} = \gamma_{p0} + \sum_{q=1}^{Q} \gamma_{pq} Z_{qj} + \mu_{pj} \quad (3)$$

where $\gamma_{00}$ and $\gamma_{p0}$ are estimated intercepts in the "Vehicle company & year"-unit level; $Z_{qj}$ is the $q$th independent variable for "Vehicle company & year"-unit $j$; $\mu_{0j}$ and $\mu_{pj}$ are the random effects varying across "Vehicle company & year"-units for the crash-level intercept and covariate $p$, and they are assumed as normal distributions with means zero and variances $\sigma_0^2$ and $\sigma_k^2$, respectively.

Both $\beta_{0j}$ and $\beta_{pj}$ vary with the different "Vehicle company & year" units, in which two components are combined to decide the coefficient values. First, it's assumed that they have linear relationships with the level 2 covariates $Z_{qj}$, because various environmental, road, and vehicle features may result in different severity levels or crash types. Second, besides the fixed parts which depend on the level 2 covariates $Z_{qj}$, random effects are also included ($\mu_{0j}$ and $\mu_{pj}$). The random effects between "Vehicle company & year" units only vary across the different units, but in the same unit, they are constant for the crash. Previous studies have shown that considering these random effects, potential random variations across "Vehicle company & year" units are allowed and correlations within them can be explained (Jones and Jorgensen, 2003; Kim et al., 2007).

The full model with Equation (4) is the hierarchical model with both random intercept and random slope (Huang et al., 2008).

$$logit(\pi_{ij}) = \log\left(\frac{\pi_{ij}}{1-\pi_{ij}}\right)$$
$$= \gamma_{00} + \sum_{q=1}^{Q} \gamma_{0q} Z_{qj} + \mu_{0j} + \sum_{p=1}^{P} \gamma_{p0} X_{pij} + \sum_{p=1}^{P}\sum_{q=1}^{Q} \gamma_{pq} Z_{qj} X_{pij}$$
$$+ \sum_{p=1}^{P} \mu_{pj} X_{pij} + \varepsilon_{ij} \quad (4)$$

**2.2.2 Bayesian inference**

To calibrate the hierarchical model, this study employed Bayesian inference. Bayesian inference technique is a prevailing way to explicitly model the hierarchical structure, in which, prior beliefs and the likelihood function of data at hand are fused to obtain the marginal posterior distribution. The distinctions between fixed and random effects disappear since all effects are now considered to be random and the hierarchical structure is accounted for. Compared to classical frequentist methods, Bayesian inference shows a lot of theoretical and practical advantages in road safety analysis, such as the good capability to handle small size data and deal with missing data commonly occurred in crash records, allowing comparison of any number of non-nested models, and considering hierarchies in the model (Mitra and Washington, 2007; Silva and Filho, 2020; Kia et al. 2021).

Prior distribution is a material part in Bayesian inference, and the following three kinds of prior distribution are commonly used (Olia, 2018): (a) Strong informative prior distributions based on expert knowledge or previous investigation; (b) Weak informative prior distributions that do not dictate the posterior distribution significantly but are able to prevent inappropriate inferences; (c) Uniform priors that could interpret evidence from the data probabilistically. (Ahmed et al., 2018; Dong et al., 2014; El-Basyouny and Sayed, 2009; Wali et al., 2018; Boggs et al., 2020). In the absence of prior information, uniform priors were used in this study. For the regression coefficients ($\gamma_{00}, \gamma_{0q}, \gamma_{p0},$ and $\gamma_{pq}$), normal distributions (0, 1000) were assumed. A study of Fink (1997) showed that the conditional conjugacy property of inverse-gamma priors suggested more flexible mathematical properties, so the variance $\sigma_0^2$ and $\sigma_k^2$ were assumed to be distributed as Gamma (0.001, 0.001).

A well-known computing approach for Bayesian inference, the Markov chain Monte Carlo (MCMC) method (Gelman, 2006), was used in this study to better approximate the target posterior distribution. Two parallel MCMC chains were initiated for each model, and 5,000 starting iterations in each chain were dropped as burn-in, whereas 10,000 iterations in each chain were used for generating the descriptive statistics for posterior estimates. In summary, the posterior estimates were based on 20,000 MCMC iterations (10,000 in each of the chains). The MCMC chains were reasonably converged, since the ratio of pooled- and within-chain interval widths were around 1. (Gelman et al., 2013).

For better interpreting the results of the hierarchical Bayesian models, the odds ratios were calculated (Odds ratio =$e^\gamma$) to show relative likelihoods. For example, in the analysis of crash severity, with the independent variable switching from 0 to 1, the odds of high severity crash increases/decreases by a value of |$e^\gamma$-1|. This study used the 95% Bayesian credible interval (95% BCI) to examine the significance of variables. The coefficient estimations were identified to be significant if the 95% BCIs didn't cover "0" or the 95% BCIs of odds ratio didn't cover "1".

In this study, Watanabe-Akaike Information Criterion (WAIC) (Watanabe, 2010) and Leave-one-out cross-validation (LOO) (Vehtari et al., 2017) were used to measure the model performance and select the best fitting model. LOO and WAIC have various advantages over simpler estimates of predictive error. WAIC can be viewed as an improvement on the deviance information criterion (DIC). It has been known that DIC has some problems for Bayesian models, which arises in part from not being fully Bayesian where DIC is based on a point estimate (Plummer, 2008). Unlike DIC, WAIC is fully Bayesian in that it uses the entire posterior distribution, and it is asymptotically equal to Bayesian cross-validation (Vehtari et al., 2015). However, the study of Vehtari et al. showed that LOO was recommended to be tried in the finite case with influential observations. Thus, in our study, the best-fitting models were those with the lowest WAIC and LOO values.

The hierarchical Bayesian approach was performed to model the binary outcome using the "brm" package in the statistical software R (version 3.4.4).

## 3. Results and Discussion

### 3.1 Impacting factors on crash severity

As shown in Figure 2, in the conventional driving mode, the proportion of crashes with injury was 18%, similar to that of the autonomous driving mode (22%).

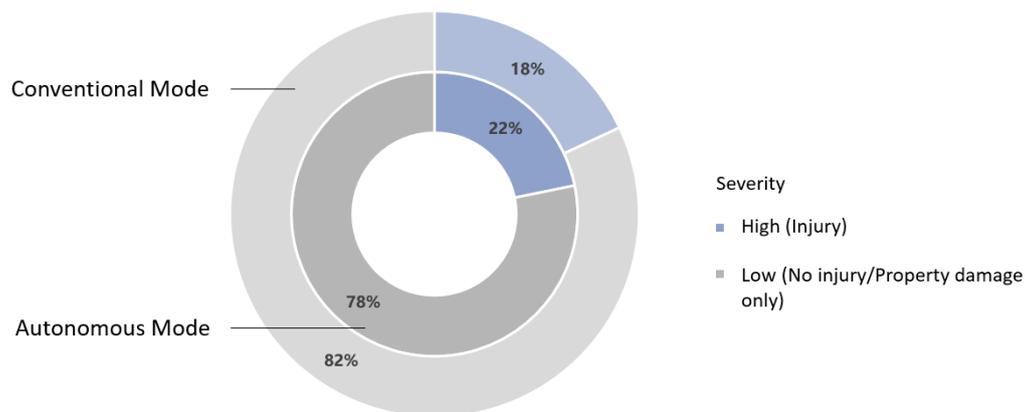

**Fig. 2. Statistical results of crash severity in both driving modes**

The best-fit hierarchical Bayesian models with the lowest WAIC and LOO for crash severity in both driving modes were finally selected. The model for the autonomous mode included a total of eight explanatory variables, and that for the conventional mode contained seven explanatory variables. All these included variables were statistically significant, and removing any of them would reduce the systemic utility of these models. Tables 3 and 4 presents the results of the two hierarchical Bayesian models. To represent the data more intuitively, odds ratios (OR) are plotted in Figures 3 and 4.

For the autonomous mode, raining presence, mixed land use, muni line presence, bike lanes presence, two sidewalks presence, and moving vehicle state were all positively associated with the crash severity, whereas daytime, and daily visitors' flowrate (DVF) less than 3418 person-times had negative effects on crash severity.

For the conventional mode, the number of lanes at the crash site, bike lanes presence, turning movement presence, moving, and vehicle state all positively affected crash severity, whereas DVFs (<3418, 3418~11982, and 11982~40040 person-times) were negatively associated with crash severity.

**Tab. 3. The hierarchical Bayesian model for crash severity in the autonomous mode**

| Parameters | Estimate (std error) | Odds ratio (95% confidence interval) |
|---|---|---|
| **Fixed effects** | | |
| **Environmental variables** | | |
| Daytime | -0.23 (0.08) | 0.79 (0.66~0.96) |
| Night* | 0 | 1 |
| Daily visitors' flowrate (DVF)< 3418 person-times | -0.16 (0.10) | 0.85 (0.76~0.95) |
| DVF>40040 person-times * | 0 | 1 |
| Raining presence | 0.09 (0.27) | 1.09 (1.03~1.16) |
| Raining absence* | 0 | 1 |
| Mixed land use# | 0.17 (0.12) | 1.19 (1.02~1.38) |
| Residential land use * | 0 | 1 |
| Muni line presence | 0.39 (0.09) | 1.48 (1.06~2.05) |
| Muni line absence* | 0 | 1 |
| **Road variables** | | |
| Bike lanes presence | 0.20 (0.09) | 1.22 (1.08~1.38) |
| Bike lanes absence* | 0 | 1 |
| Two sidewalks presence | 0.24 (0.17) | 1.27 (1.03~1.57) |
| Absence or only one sidewalk* | 0 | 1 |
| **Vehicle variables** | | |
| Vehicle state-moving | 0.45 (0.28) | 1.57 (1.13~2.18) |
| Vehicle state-stopped* | 0 | 1 |
| **Intercept (level 1)** | 0.45 (0.32) | 1.57 (1.01~2.44) |
| **Random effects** | | |
| Vehicle state-moving | 0.16 (0.13) | 1.17 (1.00~1.38) |
| Intercept (Vehicle company & year) | 0.09 (0.10) | 1.09 (1.06~1.13) |
| **WAIC** | 62.8 | |
| **LOO** | 63.4 | |

*Note: * denotes the reference group; # denotes the Random variable*

**Tab. 4. The hierarchical Bayesian model for crash severity in the conventional mode**

| Parameters | Estimate (std error) | Odds ratio (95% confidence interval) |
|---|---|---|
| **Fixed effects** | | |
| **Environmental variables** | | |
| Daily visitors' flowrate (DVF)< 3418 person-times[#] | -1.01 (0.28) | 0.36 (0.21~0.64) |
| 3418<DVF<11982 person-times | -0.96 (0.22) | 0.38 (0.25~0.59) |
| 11982<DVF<40040 person-times | -0.89 (0.21) | 0.41 (0.27~0.61) |
| DVF>40040 person-times * | 0 | 1 |
| **Road variables** | | |
| Number of lanes at crash site>2 | 0.17 (0.10) | 1.19 (1.01~1.40) |
| Number of lanes at crash site≤2* | 0 | 1 |
| Bike lanes presence | 0.35 (0.09) | 1.42 (1.19~1.70) |
| Bike lanes absence * | 0 | 1 |
| **Vehicle variables** | | |
| Turning movement presence | 0.20 (0.10) | 1.22 (1.02~1.51) |
| Turning movement absence * | 0 | 1 |
| Vehicle state-moving | 0.22 (0.11) | 1.25 (1.02~1.57) |
| Vehicle state-stopped* | 0 | 1 |
| **Intercept (level 1)** | 0.74 (0.23) | 2.09 (1.32~3.19) |
| **Random effects** | | |
| DVF<3418 person-times | 0.30 (0.23) | 1.35 (1.03~2.53) |
| Intercept (Vehicle company & year) | 0.21 (0.28) | 1.23 (1.01~3.63) |
| **WAIC** | 52.6 | |
| **LOO** | 53.3 | |

*Note: * denotes the reference group; [#] denotes the Random variable*

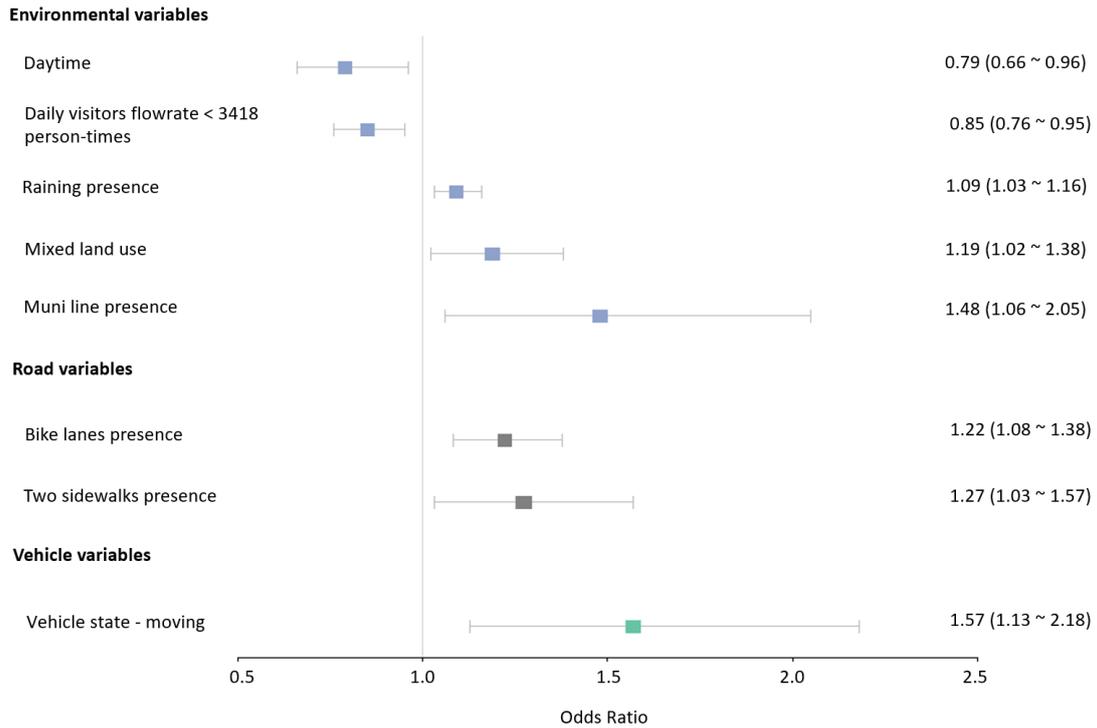

Fig. 3. Odds Ratio of the influencing factors for crash severity in the autonomous mode

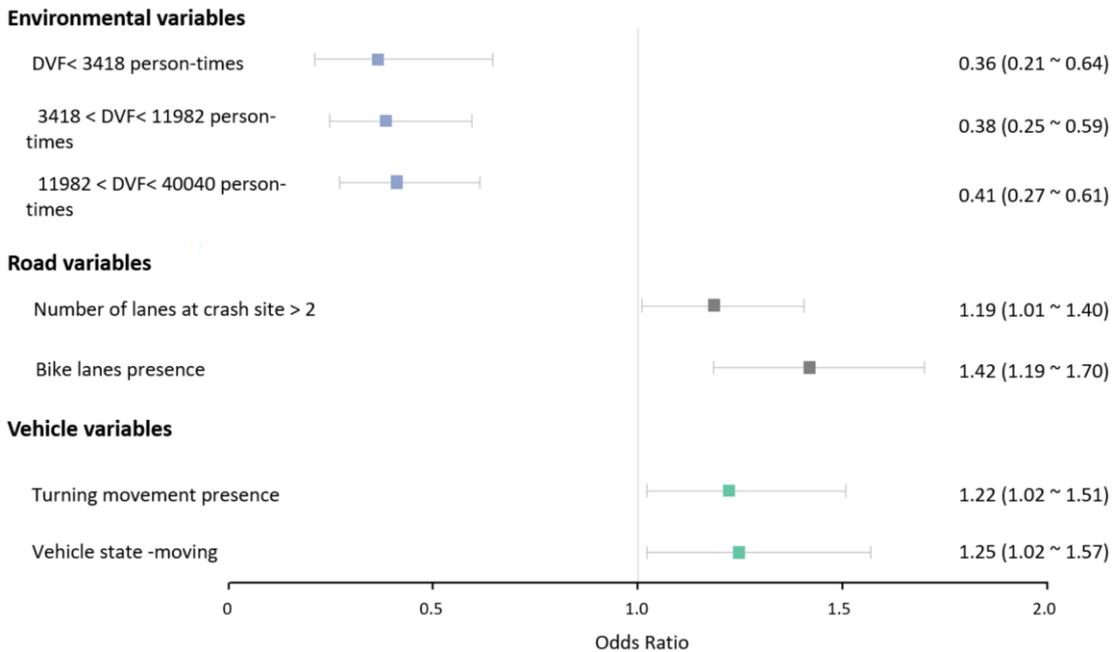

Fig. 4. Odds Ratio of the influencing factors for crash severity in the conventional mode

Detailed explanations of these influencing variables are provided below from the following three aspects. Additionally, Figure 5 demonstrates the comparison of the same influencing factors for injury crash propensity in the autonomous driving mode and conventional driving mode. In Figure 5, the posterior distribution of each influencing factor's OR is plotted.

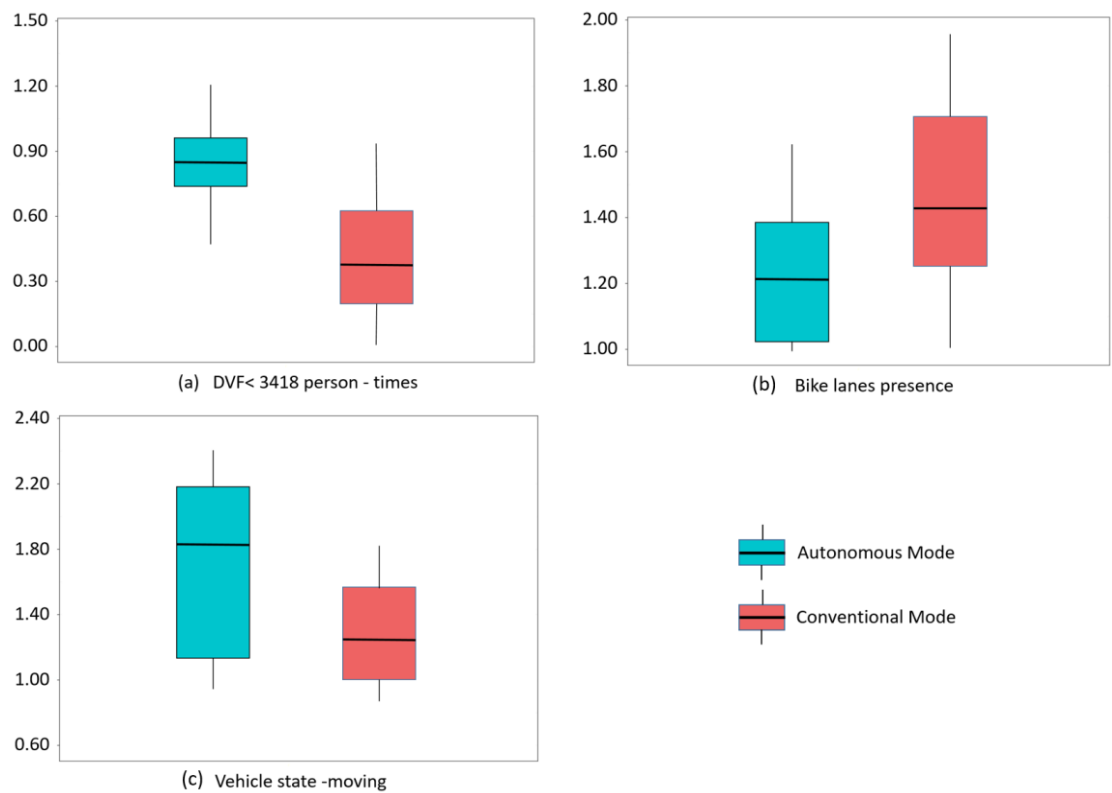

**Fig. 5. Comparison of the same influencing factors for crash severity in autonomous and conventional driving modes**

### 3.1.1 Environmental variables

The results revealed a positive statistically significant correlation between the population flow size and crash severity in both driving modes. Compared to the maximum level of DVF (>11982 person-times), the likelihood of high severity decreased in the other three levels in the conventional mode by 64%, 62%, and 59% respectively. (DVF<3418 person-times, OR=0.36; DVF 3418~11982 person-times, OR=0.38; DVF 11982~40040 person-times, OR=0.41) (See Table 4 and Figure 4). As shown in Table 3 and Figure 3, when compared to the maximum level of DVF, there was a decrease of 15% in the likelihood of high crash severity if DVF was less than 3418 person-times in the autonomous driving mode (OR=0.85). The effect of other levels of DVF in this mode was not statistically significant. Figure 5 (a) showed the different impacts of the minimum level of DVF on injury crash propensity in different driving modes.

As the results showed, injury crashes were more likely to occur in areas with higher DVF. There are often great safety hazards in densely populated areas such as bus stations, schools, and residential areas (Chen et al., 2020). For the conventional driving mode, passing through these areas, nervousness, and anxiety of drivers may increase, which would affect the driving behavior (Chen, 2015). For the autonomous driving mode, crowded areas would significantly increase the conflict points between vehicles, other transportation, and pedestrians, and the perceived ability of AVs may be influenced. However, AVs can reduce crash severity by avoiding driver errors (e.g., speeding, fatigue driving, aggressive driving, distracted driving, slow reaction times, etc.). Compared with humans, AVs are supposed to have better deceleration performance, leading to low collision speed in crowed sites (with higher DVF). Therefore, the impact of DVF on crash severity in the autonomous driving mode is smaller than that in the conventional driving mode.

Compared with the conventional driving mode, several other factors also significantly affected the likelihood of high crash severity in the autonomous mode (See Table 3 and Figure 3). The presence of muni lines increased the likelihood of high crash severity by 48% (OR=1.48). Raining increased the likelihood of high crash severity in the crash by 9% (OR=1.09). Compared to residential land use, there was a 19% growth in the possibility of high crash severity in mixed land-use types (OR=1.19). Relative to the nighttime, daytime would reduce the likelihood of high crash severity in the crash by 21% (OR=0.79).

The existence of the muni line may lead to a more complicated traffic environment in which buses, conventional vehicles, and AVs are mixed. As mentioned before, AVs may lead to some problems in the mixed flow. Previous studies suggest that AVs should be combined with Public Transportation (PT) systems to reduce labor costs, expand service hours and optimize the spatial and temporal allocation of the PT services (Shen et al., 2018; Lam et al., 2016). However, the safety issues caused by the combination of AVs and PT should not be ignored in the current stage of AVs. In addition, mixed land-use patterns typically exhibit diverse land-use types leading to complex roadway layouts, line of sight occlusion, and other problems. A study also found that AVs may have critical issues in the case of complex environments (Aria et al., 2016). Human drivers have a certain understanding of the general structure of each kind of thing on the road, and they can rapidly imagine the shape of the occluded object and deal with the occlusion problems well. But for AVs, this is a problem that may lead to untimely braking. At present, there are still problems in the total factor recognition and perception of current autonomous driving vehicles in complex environments, one of which is that autonomous driving vehicles cannot effectively and timely identify all the factors that may affect driving safety.

The camera system on AVs depends on the brightness of the scene to determine the intensity of image pixels (Vargas et al., 2021). Since night vision images have fewer texture details and low contrast, dim light in the evening would reduce the ability of AVs to recognize the surrounding scene and lack sufficient reaction time to avoid serious crashes. (Jaarsma and De Vries, 2014; Owens et al., 2019). Besides, rain may obscure the edges of objects, making them difficult to recognizable. Although radar may not be affected by dark conditions or rain, vulnerable road users cannot be identified accurately (Owens et al., 2018). Human drivers also have poor visualization at night, however, most of them would be more vigilant, which may lead to relatively low speeds to avoid serious crashes. Visibility may provide safety improvement with the advance of machine vision and digital image processing techniques. In addition, roadway, intersection, and personal lighting, reflective materials, night vision, and educational interventions are also important (Chiou et al., 2013).

### 3.1.2 Road variables

The existence of bicycle lanes increased the likelihood of the high crash severity compared to the road with only motor lanes in both driving modes. The likelihood increased by 22% in the autonomous driving mode (OR=1.22), while in the conventional mode, there was an increase of 42% (OR=1.42), representing a greater positive impact (See Table3, Table 4, and Figure 5 (b)). The existence of bicycle lanes means there are more non-motor vehicles in these sections. Compared with the collision between vehicles, non-motor vehicles are more frequently injured in crashes. AVs' technologically advanced sensors and algorithms, and the potential ability of bicycles to communicate with AVs via transponders are viewed as reasons for greater ability to perceive cyclists (Pettigrew et al., 2020). Thus, human drivers are more vulnerable to the emergence of bicycles on the road.

As shown in Table 3 and Figure 3, compared to no crosswalk roadway or road with only one side crosswalk, the likelihood of high crash severity was increased at roadways with crosswalks on both sides by 27% in the autonomous driving mode (OR=1.27). Lots of research about the interaction between pedestrians and AVs have shown that the ability of AVs to detect and understand responses from pedestrians and respond appropriately is not yet complete (Jayaraman et al., 2018; Utriainen, 2020). Without pedestrian-to-driver communication (e.g., eye contact), pedestrian behavior becomes more unpredictable (Straub and Schaefer, 2019).

As for the conventional driving mode, compared with the number of lanes at the crash site less than 2, the possibility of high crash severity raised by 19% (OR=1.19) if the number of lanes was more than 2 (See Table 4 and Figure 4). As with previous studies, more lanes are positively related to higher crash severity, since they commonly mean higher speed, which may cause higher crash severity (Li et al., 2013; Aziz et al., 2013). The advantages of AVs, such as avoiding driver errors and better deceleration performance, make them perform better than human drivers in the multi-lane scenario.

### 3.1.3 Vehicle variables

The state of the vehicle impacted the likelihood of the high crash severity in both driving modes. When compared to stopped vehicles, the possibility of the high crash severity rose by 25% with moving vehicles in the conventional mode (OR=1.25). In the autonomous mode, there was a 57% increase (OR=1.57) (See Table 3, Table 4, and Figure 5 (c)). From the perspective of kinetic energy, it is easy to explain that moving vehicles would cause more serious consequences. However, the different degrees of increase in the two modes are worthier attention. Previous studies have reported that AVs have a significant impact on the uncertainty, conflict, and stability of mixed traffic, which are highly associated with the severity of crashes (Zheng et al., 2020; Mahdinia et al., 2021). Although the safety performance could substantially improve with a high penetration of AVs, AVs would adversely affect the traffic environment at lower penetration (Arvin et al., 2019). Besides, drivers may game with the limitations of AVs and behave more aggressively in their vicinity (Arvin et al., 2020). Therefore, popularizing AV knowledge for conventional drivers and improving their vigilance in mixed traffic may help reduce crash severity. While studying car-following in mixed traffic, the interaction between human drivers and AVs should be paid more attention

(Liu et al., 2021).

As shown in Table 4 and Figure 4, compared with vehicles operating straight ahead, the likelihood of the high crash severity rose by 22% (OR=1.22) when the vehicle was making a turning motion in the conventional mode. Turning movements are usually associated with much higher attention than normal moving (Yu et al., 2019c). AVs with powerful perception and decision-making systems may perform better than human drivers.

**3.2 Impacting factors on crash type**

As shown in Figure 6, the proportion of rear-end collision was 40% under the conventional driving model, while a much higher likelihood of AVs involved rear-end crashes was found to be 59%. This 19-percent difference in the rear-end crash frequencies may indicate the significant different driving characteristics between AVs and conventional vehicles.

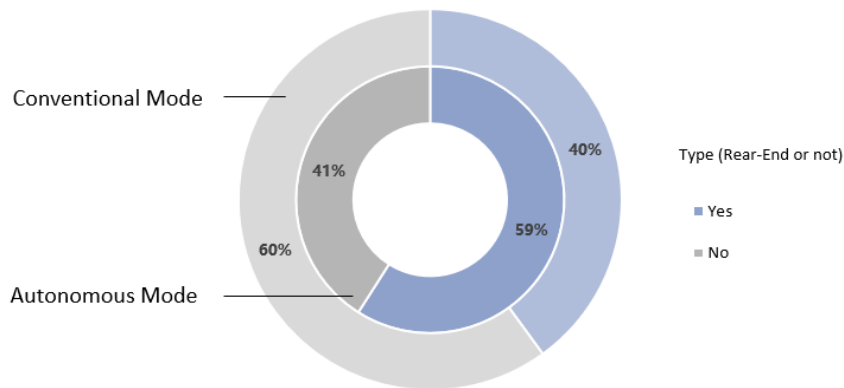

**Fig. 6. Statistical results of crash type in both driving mode**

The best-fit hierarchical Bayesian models for crash type in both driving modes were selected, according to the lowest WAIC and LOO. There were eight explanatory variables in the crash type model under the autonomous mode and six explanatory variables under the conventional mode. Each included variable was statistically significant. Tables 5 and 6 present the results of the two hierarchical Bayesian models, while Figures 7 and 8 illustrate OR for representing the data more intuitively.

Under the autonomous mode, no non-motor vehicles and pedestrians involved in the crash, on-street parking, number of driveways less than 4, divided median, deceleration presence, and turning movement had positive associations with the likelihood of rear-end crashes, whereas trees and disengagement presence were negatively associated with that.

As for the conventional driving mode, the slope of the road more than 3, divided median, and turning movement were positively associated with the rear-end likelihood, whereas DVF less than 3418 person-times, dry surface, and the presence of bike lines had negative association with the probability of rear-end crashes.

**Tab. 5. The hierarchical Bayesian model for crash type in the autonomous mode**

| Parameters | Estimate (std error) | Odds ratio (95% confidence interval) |
|---|---|---|
| **Fixed effects** | | |
| **Environmental variables** | | |
| Trees presence | -0.21 (0.19) | 0.81 (0.68~0.96) |
| Trees absence* | 0 | 1 |
| No involved (non-motor vehicles and pedestrians)# | 0.12 (0.12) | 1.13 (1.03~1.23) |
| Others involved* | 0 | 1 |
| On street parking presence | 0.38 (0.10) | 1.46 (1.09~1.95) |
| On street parking absence* | 0 | 1 |
| **Road variables** | | |
| Number of driveways<4 | 0.13 (0.11) | 1.14 (1.05~1.23) |
| Number of driveways≥4* | 0 | 1 |
| Divided median presence | 0.36 (0.20) | 1.43 (1.02~2.01) |
| Divided median absence* | 0 | 1 |
| **Vehicle variables** | | |
| Disengagement presence | -0.15 (0.14) | 0.86 (0.75~0.99) |
| Disengagement absence* | 0 | 1 |
| Deceleration presence# | 0.4 (0.28) | 1.49 (1.03~2.23) |
| Deceleration absence* | 0 | 1 |
| Turning movement presence | 0.51 (0.39) | 1.67 (1.01~2.75) |
| Turning movement absence* | 0 | 1 |
| **Intercept (level 1)** | -0.35 (0.29) | 0.70 (0.52~0.96) |
| **Random effects** | | |
| No involved (non-motor vehicles and pedestrians) | 0.07 (0.07) | 1.07 (1.00~1.15) |
| Deceleration presence | 0.13 (0.09) | 1.14 (1.02~1.27) |
| Intercept (Vehicle company & year) | 0.22 (0.17) | 1.25 (1.11~1.40) |
| **WAIC** | 70.2 | |
| **LOO** | 72.1 | |

*Note: * denotes reference group; # denotes Random variable*

**Tab. 6. The hierarchical Bayesian model for crash type in the conventional mode**

| Parameters | Estimate (std error) | Odds ratio (95% confidence interval) |
|---|---|---|
| **Fixed effects** | | |
| **Environmental variables** | | |
|   Daily visitors' flowrate (DVF)< 3418 person-times | -0.38 (0.11) | 0.68 (0.55~0.86) |
|   DVF>40040 person-times * | 0 | 1 |
|   Dry surface | -0.33 (0.14) | 0.72 (0.55~0.95) |
|   Wet surface* | 0 | 1 |
| **Road variables** | | |
|   Bike lanes presence | -0.28 (0.13) | 0.76 (0.58~0.98) |
|   Bike lanes absence* | 0 | 1 |
|   Slop>3 | 0.24 (0.11) | 1.27 (1.02~1.55) |
|   Slop≤3* | 0 | 1 |
|   Divided median presence# | 0.38 (0.17) | 1.46 (1.03~2.01) |
|   Divided median absence* | 0 | 1 |
| **Vehicle variables** | | |
|   Turning movement presence | 0.26 (0.12) | 1.30 (1.00~1.65) |
|   Turning movement absence* | 0 | 1 |
| **Intercept (level 1)** | 0.26 (0.16) | 1.30 (0.89~1.72) |
| **Random effects** | | |
|   Divided median presence | 0.21 (0.18) | 1.23 (1.03~1.48) |
|   Intercept (Vehicle company & year) | 0.20 (0.14) | 1.22 (1.02~1.77) |
| **WAIC** | 94.2 | |
| **LOO** | 94.6 | |

*Note: * denotes reference group; # denotes Random variable*

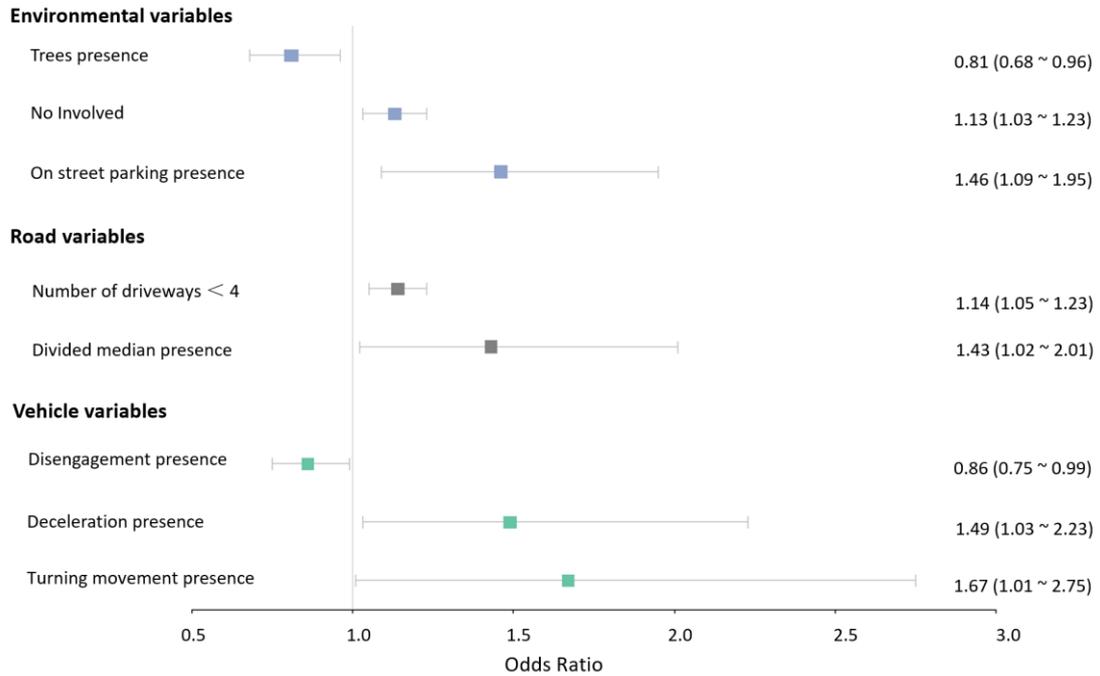

Fig. 7. Odds Ratio of the influencing factors for crash type in the autonomous Mode

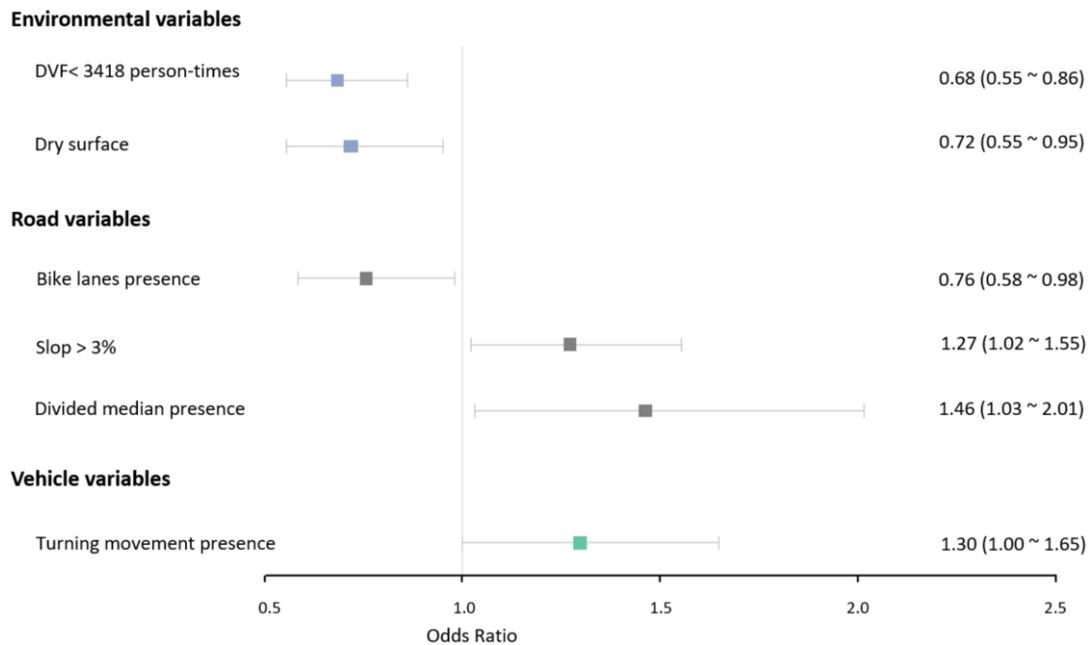

Fig. 8. Odds Ratio of the influencing factors for crash type in the conventional mode

A full discussion of these influencing variables is provided below from the following three aspects. Figure 9 demonstrates the comparison of the same influencing factors for crash type in the autonomous driving mode and conventional driving mode. The posterior distribution of each influencing factor's OR is plotted in Figure 9.

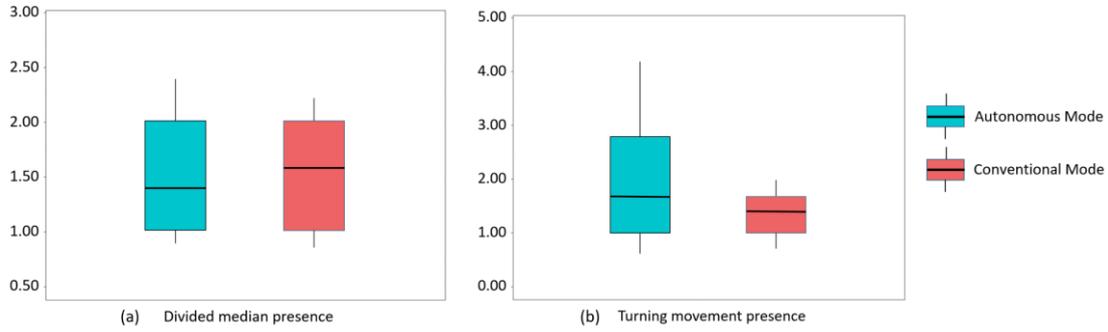

**Fig. 9. Comparison of the same influencing factors for crash type in the autonomous driving mode and conventional driving mode**

### 3.2.1 Environmental variables

As shown in Table 5 and Figure 7, in terms of the presence of trees alongside the road, the likelihoods of rear-end crashes had a decrease of 19% in the autonomous mode (OR=0.81). In particular, compared to someone involved in the crash (including non-motor vehicles and pedestrians), there was an upward trend in the possibility of rear-end crashes of 13% (OR=1.13) when no non-motor vehicles and pedestrians were involved. The existence of on-street parking raised the likelihood of rear-end crashes by 46% (OR=1.46).

Trees would help establish a more clearly defined broadside boundary and create more enclosed streetscapes along streets (Dumbaugh and Gattis, 2005; Harvey and Aultman-Hall, 2015), which can help vehicles to maintain low speed. Besides that, the presence of trees is conducive to calm traffic and can help AVs and manual drivers to follow each other harmoniously in mixed traffic conditions (Van Treese II et al., 2017). Therefore, the occurrence of rear-end crashes would decrease. There is no doubt that the overall harmony of traffic flow may help reduce nearly all types of crashes. However, comparing with rear-end crashes, the impact on other types is smaller. For example, sideswipe is always accompanied by lane changes, and factors between adjacent lanes have a greater impact (Lee et al., 2006). As for single-vehicle accidents (e.g., hit an object, hit other transportation, etc.), they are mainly affected by exogenous factors (e.g., alcohol use and road geometry), rather than variables about traffic flow (Christoforou et al., 2011). Besides, some studies have proposed reasonable methods for arranging trees to make the driving environment safer, such as selectively planting small-stature trees, removing trees in high-risk locations, etc. (Harvey and Aultman-Hall, 2015; Mok et al., 2006; Roodt, 2012). A crash related to pedestrians or non-motor vehicles may not involve another vehicle. As a result, the proportion of rear-end crashes is low.

The results of this study showed a positive correlation between rear-end crashes and on-street parking, which was consistent with other studies (Das and Abdel-Aty, 2011). On-street parking complicates the road environment, and the lack of visibility due to parked cars makes drivers or AVs unaware of oncoming pedestrians or other vehicles entering into the roads which would lead to emergency brakes (Gitelman et al., 2012). However, human drivers would render extra consciousness while driving through on-street parking areas (Biswas et al., 2017) and may better adapt to the scene. Some possible solutions provided by other studies may decrease rear-end crashes, such as prohibiting on-street parking near some specific locations like a designated pedestrian crossing, intersection, etc., and setting them back 23m or more from the yield line (Chen et al., 2017; Biswas et al., 2017).

Compared with the autonomous driving mode, several other factors also influenced the rear-end crash propensity in the conventional driving mode (See Table 6 and Figure 8). When compared to the maximum level of DVF (>11982 person-times), the likelihood of rear-end crashes decreased by 32% (OR=0.68) if the DVF was less than 3418 person-times in the conventional driving mode. Compared to the wet surface, the likelihood of rear-end crashes dropped by 28% (OR=0.72) with the dry surface.

As mentioned before, passing through areas with large DVF, nervousness and anxiety of drivers may increase, which would affect the driving behavior (Chen, 2015). However, AVs would be less influenced by DVF because of their better deceleration performance and the ability to avoid driver errors. Water film between tires and road surface on the wet road may reduce the adhesion coefficient of the road surface, which may increase the braking distance of vehicles (Yan et al., 2005). Vehicle braking, steering, and other performance would also be affected. When the front car brakes suddenly, the rear car may not stop in time, which is called a water slide. Considering the better performance of AVs in braking, dry or wet road conditions have little effect on them. Few AVs involved collisions happened on the wet road surface (less than 3%). It may be related to the fact that most drivers choose the conventional driving mode on rainy or snowy days.

### 3.2.2 Road variables

The existence of a divided median increased the likelihood of the rear-end crashes compared to the road without it in both driving modes. The odds rose by 43% (OR=1.43) in the autonomous driving mode, while in the conventional driving mode, there was a 46% increase (OR=1.46) (See Table 5, Table 6, and Figure 9 (a)). The divided median is used to prevent cross-median crashes, but it may encourage a major increase in speed among drivers (Afghari et al., 2018; Hu and Donnell, 2010). Compared with other types, high operating speeds have a more significant impact on rear-end crashes (Doecke et al., 2018). In addition, the complex movements at median openings would increase the risk of collision (Mohanty et al., 2021).

As for the autonomous driving mode, when compared to roads with more than 4 lanes, the likelihood of rear-end crashes increased on roads of less than or equal to 4 lanes by 14% (OR=1.14). The increase in the number of lanes would lead to an increase in almost all types of crashes (Kim et al., 2006). However, in the mixed traffic, narrow roads may exacerbate the car following problems between AVs and conventional vehicles, leading to an increase in rear-end crashes. Many studies have presented that the implementation of dedicated lanes for AVs may improve the safety and efficiency of the traffic and the benefits of AVs can be maximized (Kockelman et al., 2016; Milakis et al., 2017; Schoenmakers et al., 2021), but this method is difficult to implement on roads with few lanes.

Compared to the road with a slope less than or equal to 3%, the likelihood of rear-end crashes substantially rose at roadways with a slope more than 3% by 27% in the conventional driving mode (OR=1.27) (See Table 6 and Figure 8). Bike lanes decreased the likelihood of rear-end crashes in the conventional mode by 24% (OR=0.76).

It might be because that in the conventional driving mode, under the combined action of gravity and inertia force, when the longitudinal slope of the road is relatively high, vehicles would produce a large acceleration insensibly and their speed continues to increase. If vehicles are unable to slow down in time under special circumstances, high driving speed may result in a rear-end collision. The autonomous driving system has better performance in the control of acceleration and speed. As for bike lanes, some studies show that they could reduce vehicular speeds and conflicts by separating bicyclists from vehicles with bicyclists' designated path (Chen et al., 2012; Park et al., 2015). Besides, they help make bicyclists and human drivers more vigilant (Sadek et al., 2007). Human drivers are more vulnerable to the emergence of bicycles on the road.

**3.2.3 Vehicle variables**

The movement of the vehicles impacted the likelihood of rear-end crashes in both driving modes. When compared to proceeding straight, there was a 30% increase in the odds of the rear-end crashes with vehicles turning left or right in the conventional mode (OR=1.30). In the autonomous mode, turning movement's impact was more significant with an increase of 67% (OR=1.67) (See Table 5, Table 6, and Figure 9 (b)). Rear-end collisions are often caused by the deceleration of the vehicle in front and the ineffective reaction of the following driver, which would occur more frequently when there is turning movement (Wang et al., 2003; Wang, 2006). As mentioned before, turning at the mixed flow is a challenging task for AVs. Current studies also observed that AVs are more likely to be rear-ended when turning (Boggs et al., 2020).

Compared with the conventional driving mode, another two factors also influenced the rear-end crash propensity in the autonomous driving mode. As shown in Table 5 and Figure 7, there was a decrease of 14% in the possibility of rear-end crashes with the occurrence of disengagement in the autonomous mode (OR=0.86). Besides, the presence of deceleration increased the likelihood of rear-end crashes by 49% (OR=1.49). It has been confirmed that AVs in the autonomous driving mode were more likely to get involved in rear-end crashes than disengaged AVs (Boggs et al., 2020) and most of the rear-end crashes occurred with the AV hit from the rear by an upcoming vehicle. The possible reason might be the incompatible reaction time between AVs and human drivers (Favarò et al., 2017). AVs have proven to be effective in avoiding collisions with the vehicles in front, however, the following human drivers may not react opportunely with the sudden deceleration of AVs (Dixit et al., 2016; Sinha et al., 2021). Consequently, it was necessary to enhance AVs' ability to react to rear-end crashes by considering the different reaction capability of manually operated vehicles and AVs.

**3.3 Model comparison**

To examine whether the hierarchical Bayesian models with random intercept and random slopes are superior, this study made a comparison among models with different structures. Overall, models in both driving modes can be grouped into the following categories: (a) Bayesian logistic regression models (with only fixed effects) (b) hierarchical Bayesian models with random intercept (c) hierarchical Bayesian models with both random intercept and random slopes.

The lower the WAIC or LOO value, the better the model. As shown in Table 7, hierarchical Bayesian models with both random intercept and random slopes performed better than the other two kinds of models with lower WAIC and LOO.

**Tab. 7. WAIC and LOO of hierarchical Bayesian models with different structures**

|  | Bayesian logistic regression models (with only fixed effects) | | Hierarchical Bayesian models with random intercept | | Hierarchical Bayesian models with both random intercept and random slopes | |
|---|---|---|---|---|---|---|
|  | WAIC | LOO | WAIC | LOO | WAIC | LOO |
| Models for crash severity in the autonomous mode | 74.4 | 76.5 | 64.5 | 64.9 | 62.8 | 63.4 |
| Models for crash severity in the conventional mode | 60.9 | 61.9 | 53.3 | 53.7 | 52.6 | 53.3 |
| Models for crash type in the autonomous mode | 84.5 | 85.9 | 71.4 | 73.6 | 70.2 | 72.1 |
| Models for crash type in the conventional mode | 102.2 | 103.1 | 95.7 | 96.1 | 94.2 | 94.6 |

In addition, the differences between models with different observation units were also explored. As shown in Figure 10 (a), in this study, vehicles with the same vehicle year from the same company (i.e., the "Vehicle company & year" unit) were considered as a cluster (Level 2) to alleviate the effects of unobserved heterogeneity, and there were several sub-clusters per cluster, i.e., each crash (level 1). To consider the possible interconnect role of collision manner in the relationship between independent variables and crash severity, "Crash type" was taken as the criterion for classifying clusters (See Figure 10 (b)). Since studies have shown that robust parameter estimation cannot be made in the hierarchical Bayesian model if clusters in level 2 were too little (Snijders and Bosker, 2011), "Crash type" was not a dichotomous variable here, but was classified into 6 groups, including "Rear-end", "Sideswipe", "Head-on", "Hit pedestrian", "Hit non-motor vehicle", and "Others". Besides, another hierarchical Bayesian model with three levels, while considering the unobserved heterogeneities caused by both "Vehicle company & year" and "Crash type" (See Figure 10 (c)) was also attempted.

As shown in Table 8, the 2-level hierarchical Bayesian models with "Vehicle company & year" unit as level 2 had the lowest WAIC and LOO values, indicating that it performed better than the other two kinds of models.

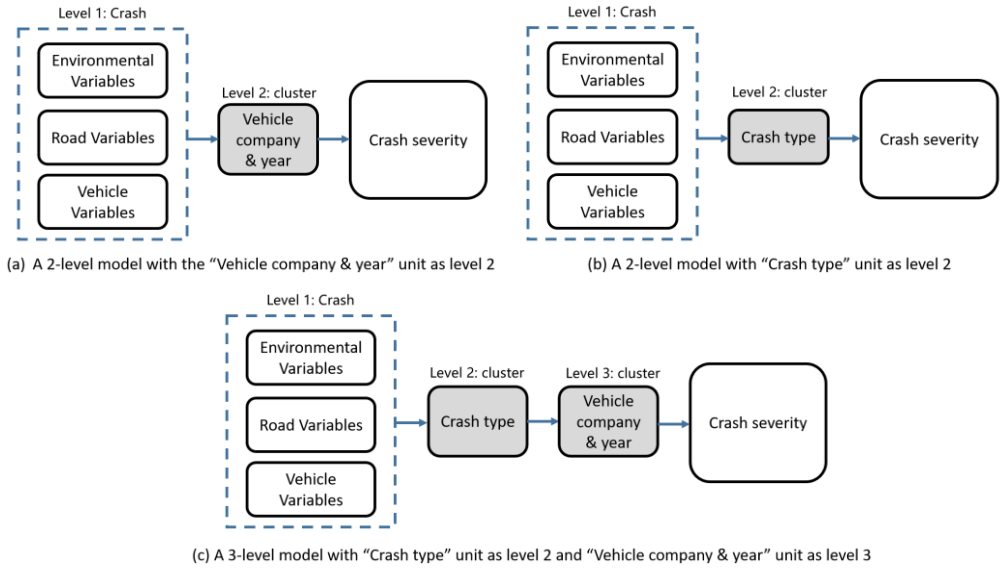

**Fig. 10. The structure of hierarchical Bayesian model**

**Tab. 8. WAIC and LOO of hierarchical Bayesian models with different observation units**

|  | The 2-level hierarchical Bayesian models with "Vehicle company & year" unit as level 2 | | The 2-level hierarchical Bayesian models with "Crash type" unit as level 2 | | The 3-level hierarchical Bayesian models with "Crash type" unit as level 2 and "Vehicle company & year" unit as level 3 | |
|---|---|---|---|---|---|---|
|  | WAIC | LOO | WAIC | LOO | WAIC | LOO |
| Models for crash severity in the autonomous mode | 62.8 | 63.4 | 65.1 | 66.0 | 63.1 | 63.8 |
| Models for crash severity in the conventional mode | 52.6 | 53.3 | 54.7 | 55.2 | 53.5 | 54.0 |

## 4. Conclusion

This study aims to analyze the divergent influences of factors on crashes under the autonomous driving mode and the conventional driving mode. By using the hierarchical Bayesian approach to consider unobserved heterogeneities, both crash type and severity were analyzed. The results showed that there were significant differences in the types and severity of influencing factors under different driving modes.

Although some influencing factors had the same positive or negative effects on crashes under both driving modes for the crash type or crash severity, their degrees were different. For example, vehicles with turning movement would decrease the likelihood of rear-end crashes in both driving modes, however, the influence under the autonomous driving mode was greater than that under the conventional driving mode. The impact of daily visitors' flowrate (DVF) on crash severity and crash type in the autonomous driving mode was smaller than that in the conventional mode since equipped with advanced sensing equipment, AVs can sense a longer distance and were superior to humans in the recognition of specific targets (e.g., face, text, etc.) (Di et al., 2017). In addition, the presence of bike lanes would lead to a great increase in the severity of crashes in the conventional driving model, while the moving vehicle state had a greater impact on the crash severity of autonomous driving. The presence of a divided median had the same positive effects for the rear-end crash type under both driving modes, but its impact is slightly higher in the conventional driving mode.

More influencing factors only had a significant impact on one of the driving modes, which was also worthy of analysis. To be specific, raining, mixed land use, the presence of the muni line, and driving at night would cause high injury severity in the autonomous mode, but in the conventional driving mode, their impacts were not significant. The problems for autonomous driving vehicles in the total factor recognition and completion of vision occlusion in complex environments may lead to this result. In addition, on-street parking would increase the likelihood of rear-end crashes in the autonomous driving mode, probably because standard road structure was easy to identify, and it was beneficial to the perception, prediction, planning, and control of AVs. The possibility of rear-end crashes would be decreased with the occurrence of disengagement, probably because of the incompatible reaction time between AVs and human drivers. More lanes would increase the possibility of high crash severity in the conventional mode because of human drivers' wrong decisions, but AVs can effectively avoid that. Besides that, AVs may have better performance in the control of acceleration and speed, since a larger slope significantly increases the possibility of rear-end crashes in the conventional driving mode, but in the autonomous mode, the influence was not significant.

The findings of this study have several practical implications:

(1) This study can help to improve driving safety in both conventional and autonomous driving modes and are beneficial to forecast crash type and severity. AVs are a complex combination of various hardware and software with high costs. Strict laws and regulations should be formulated to specify the road section and time of AVs for testing and driving (Fagnant and Kockelman, 2015). For example, the testing of AVs needs to be conducted in areas with a complex environment or at night. In addition, manufacturers should face up to the current situation that AVs are still unable to effectively deal with all factors affecting driving safety and eliminate exaggerated publicity. More importantly, despite the importance of economic benefit and efficiency for manufacturers, the improvement of the safety of AVs should not be ignored. Different influencing factors for crashes mean different perceptions and decision logic. When AVs and conventional vehicles are mixed in the road, the driving environment may be more complex, and the risk of crashes may increase. Improving the technology of autonomous driving will help to reduce the collision between them. Additionally, as for the conventional driving mode, drivers' behavior needs to be attached more importance. In a scene with more pedestrians or complex traffic flow, crashes in the conventional mode will be decreased by setting traffic signs and lines, violation penalties, and other means (Fu et al., 2018; 2019).

(2) This study shows some existing problems of autonomous driving vehicles, which are helpful to the intelligent transformation of highways and vehicles. With the popularity of the Internet of Things (IoT), vehicle information and all kinds of environmental data can be collected in real-time through the wide application of artificial intelligence and big data (Pi et al., 2020; Gerla et al., 2014). At present, the intellectualization of highways should focus more on complex road environments and solving the occlusion problem, so as to provide clearer information for vehicles. Learning ability and adaptability need to be improved for AVs, enabling them to become a moving "intelligent agent". Each "intelligent agent" can coordinate their respective routes, speeds, and distances between other vehicles to independently cope with all kinds of road conditions and unexpected situations (Gruyer et al., 2017; Watanabe and Wolf, 2018). In addition, the integration of multiple perception systems, such as visual perception systems (cameras and visual sensors), laser perception systems (laser radar), and microwave perception systems (millimeter-wave radar), will also help to reduce autonomous driving crashes.

(3) The content of this study has a certain reference value for the research of human-computer interaction. Semi-AVs will occupy the majority of the market for a long time in the future, however, it is difficult to define when the driver or vehicle should be responsible for driving. At present, AVs require human drivers to maintain control of vehicles, during the whole driving process. However, when an emergency occurs, drivers may not be able to take charge of vehicles immediately because of carelessness or gradual trust in the autonomous driving system. The warning systems shall be further improved for AVs. The core of human-computer interaction is coordination and complementarity (Yu et al., 2018; 2019a). Since autonomous driving technology has not been fully mature, semi-autonomous driving can be used as a supplement to traditional driving to reduce crashes. The human-computer conflict caused by redundant input also needs to be avoided (Flemisch et al., 2014; Wu et al., 2018).

There are some limitations in this study that should be addressed in future work. Although the Traffic Collision Involving an Autonomous Vehicle Reports (OL 316) contain much useful information, some important data, such as traffic flow, pre-crash vehicle kinematic data, driver demographics, driver perception, etc., were not recorded in the report. This study used Bayesian inference to reduce the influence of a small sample, but more crash reports will of course lead to more profound and generalizable insights. Additionally, sufficient crash data makes it possible to consider more potential relationships, such as cross-level interactions of impacting factors. Besides, it should be noted that the AV's behavior may be timid in the test stage on public roads compared to their future behavior. With the maturity and commercialization of AVs throughout the world, an updated evaluation will be required. With the AV crash data constantly updated, we will continue to add new crashes and more features to make a more detailed classification of the crash severity, and improve the accuracy and effectiveness of our models in future research.

**CRediT authorship contribution statement**

**Weixi Ren:** Conceptualization, Data curation, Methodology, Writing - original draft, Writing - review & editing. **Bo Yu**: Conceptualization, Data curation, Methodology, Writing - original draft, Writing - review & editing. **Yuren Chen**: Conceptualization, Funding acquisition, Writing - original draft. **Kun Gao**: Conceptualization, Data curation, Methodology, Writing – original. **Shan Bao**: Methodology, Writing - review & editing.

**Declaration of competing interest**

The authors declare that they have no known competing financial interests or personal relationships that could have appeared to influence the work reported in this paper.

**Acknowledgment**

This project was jointly supported by the National Natural Science Foundation of China (52102416), the Natural Science Foundation of Shanghai (22ZR1466000), the National Key Research and Development Program of China (2017YFC0803902), and Shanghai Intelligent Science and Technology Category IV Peak Discipline. We thank the State of California Department of Motor Vehicles for providing the AV crash data.

pedestrian-vehicle interactions at non-signalized intersections using vision-based trajectory data. *Transportation Research Part C: Emerging Technologies*, 105, 222-240.

Fu, T., Miranda-Moreno, L., & Saunier, N. (2018). A novel framework to evaluate pedestrian safety at non-signalized locations. *Accident Analysis & Prevention*, 111, 23-33.

Gelman, A. (2006). Prior distributions for variance parameters in hierarchical models (comment on article by Browne and Draper). *Bayesian Analysis*, 1(3), 515- 534.

Gelman, A., Stern, H.S., Carlin, J.B., Dunson, D.B., Vehtari, A., & Rubin, D.B. (2013). Bayesian Data Analysis, Third Edition, London: *Chapman and Hall/CRC*.

Gerla, M., Lee, E. K., Pau, G., & Lee, U. (2014, March). Internet of vehicles: From intelligent grid to autonomous cars and vehicular clouds. *In 2014 IEEE world forum on internet of things (WF-IoT)* (pp. 241-246). IEEE.

Gitelman, V., Balasha, D., Carmel, R., Hendel, L., & Pesahov, F. (2012). Characterization of pedestrian accidents and an examination of infrastructure measures to improve pedestrian safety in Israel. *Accident Analysis & Prevention*, *44*(1), 63-73.

Gruyer, D., Magnier, V., Hamdi, K., Claussmann, L., Orfila, O., & Rakotonirainy, A. (2017). Perception, information processing and modeling: critical stages for autonomous driving applications. *Annual Reviews in Control*, 44, 323-341.

Gucwa, M. (2014). Mobility and energy impacts of automated cars. *In Proceedings of the 2014 Automated Vehicles Symposium*, San Francisco.

Haleem, K., & Abdel-Aty, M. (2010). Examining traffic crash injury severity at unsignalized intersections. *Journal of Safety Research*, 41(4), 347-357.

Harvey, C., & Aultman-Hall, L. (2015). Urban streetscape design and crash severity. *Transportation Research Record*, *2500*(1), 1-8.

Hu, W., & Donnell, E. T. (2010). Median barrier crash severity: Some new insights. *Accident Analysis & Prevention*, *42*(6), 1697-1704.

Huang, H., Chin, H. C., & Haque, M. M. (2008). Severity of driver injury and vehicle damage in traffic crashes at intersections: a bayesian hierarchical analysis. *Accident Analysis & Prevention*, 40(1), 45-54.

Jaarsma, C. F., & De Vries, J. R. (2014). Agricultural vehicles and rural road safety: tackling a persistent problem. *Traffic injury prevention*, *15*(1), 94-101.

Jayaraman, S. K., Creech, C., Robert Jr, L. P., Tilbury, D. M., Yang, X. J., Pradhan, A. K., & Tsui, K. M. (2018, March). Trust in AV: An uncertainty reduction model of AV-pedestrian interactions. In *Companion of the 2018 ACM/IEEE international conference on human-robot interaction* (pp. 133-134).

Johnson, C., Walker, J. (2016). Peak Car Ownership: the market opportunity of electric automated mobility services. Retrieved August 1, 2021, from https://rmi.org/insight/peak-car-ownership-reportJones.

Jones, A. P., & Jorgensen, S. H. (2003). The use of multilevel models for the prediction of road accident outcomes. *Accident Analysis & Prevention*, 35(1), 59-69.

Khattak, A.J., & Wali, B. (2017). Analysis of volatility in driving regimes extracted from basic safety messages transmitted between connected vehicles.37